# Ferroelectric and anomalous quantum Hall states in bare rhombohedral trilayer graphene

Felix Winterer[1], Fabian R. Geisenhof[1], Noelia Fernandez[1,2],
Anna M. Seiler[1,2], Fan Zhang[3], R. Thomas Weitz[1,2,*]

[1] Physics of Nanosystems, Department of Physics, Ludwig-Maximilians-Universität München, Amalienstrasse 54, Munich 80799, Germany

[2]  1st Institute of Physics, Faculty of Physics, University of Göttingen, Friedrich-Hund-Platz 1, Göttingen 37077, Germany

[3]  Department of Physics, University of Texas at Dallas, Richardson, TX, 75080, USA

*Corresponding author. Email: thomas.weitz@uni-goettingen.de

**Nontrivial interacting phases can emerge in elementary materials. As a prime example, continuing advances in device quality have facilitated the observation of a variety of spontaneous quantum Hall-like states[1–8], a cascade of Stoner-like magnets[9–11], and an unconventional superconductor[9,12] in bilayer graphene. Its natural extension, rhombohedral trilayer graphene is predicted to be even more susceptible to interactions given its even flatter low-energy bands and larger winding number[13]. Theoretically, five spontaneous quantum Hall phases have been proposed to be candidate ground states[14–16]. Here, we provide transport evidence for observing four of the five competing ordered states in interaction-maximized, dually-gated, rhombohedral trilayer graphene. In particular, at vanishing but finite magnetic fields, two states with Chern numbers 3 and 6 can be stabilized at elevated and low electric fields, respectively, and both exhibit clear magnetic hysteresis. We also reveal that the quantum Hall ferromagnets of the zeroth Landau level are ferroelectrics with spontaneous layer polarizations even at zero electric field, as evidenced by electric hysteresis. Our findings exemplify the possible birth of rich interacting electron physics in a simple elementary material.**

The competition between kinetic and exchange energies in conjunction with the band topology determines the ground state properties of an interacting system. Recently, experiments featuring Chern insulator ground states in moiré superlattices without the requirement of applying any external magnetic field or magnetic dopant[8,17–21] have gained significant interest. In fact, two such quantum anomalous Hall phases were predicted to be competing ground states of the naturally occurring rhombohedral graphene systems such as AB bilayer[14,22] and ABC trilayer[14,16]. While consistent signatures have already been identified in a recent experimental examination of suspended AB bilayer graphene (BLG)[8] in principle, the susceptibility toward spontaneously gapped phases in ABC trilayer graphene (r-TLG) should be still stronger[13,16]. This is indeed the case by comparing the experimental conditions to observe the unexpectedly discovered Stoner magnetism and unusual superconductivity in lightly hole-doped BLG[9–12] and r-TLG[23,24].

In rhombohedral multilayer graphene, quasiparticles can be described by a two-component pseudospinor associated with the two low-energy sublattice sites located respectively on the two outermost layers[14,16,25–27], leading to an effective two-touching-band Hamiltonian $H = \frac{(v_F p)^N}{\gamma_1^{N-1}} \left[ \cos(N\phi)\,\sigma_x + \sin(N\phi)\,\sigma_y \right]$. (The trigonal warping effect can be appropriately ignored for this work). Here, $N$ is the number of layers, $v_F$ is the Fermi velocity in graphene, $\gamma_1$ is the nearest-neighbor interlayer hopping energy, $p$ is the momentum with $\phi = \tan^{-1} \xi p_y / p_x$, $\xi = \pm 1$ labels the $K$ and $K'$ valleys, and $\vec{\sigma}$ are the Pauli matrices acting on the layer pseudospin. A quasiparticle acquires a Berry phase of $\xi N\pi$ when encircling the band touching point at $K/K'$. The band touching nature and Berry phase quantization is protected by the sublattice or chiral symmetry and by the inversion and time-reversal symmetries[16]. Thus, a perpendicular electric field[26,27] or electron-electron interactions[14,16,27] can explicitly or spontaneously break the symmetry and open an energy gap. Moreover, rhombohedral multilayer graphene has a power-law dispersion $E \propto P^N$ that flattens with increasing $N$, as reflected by the density of states near charge neutrality $\propto E^{(2-N)/N}$ that vanishes for monolayer graphene ($N$ = 1), is constant for BLG ($N$ = 2), and diverges for thicker systems ($N > 2$). This yields the important fact[13] that the electron-electron interactions, modeled as a coupling constant in a renormalization group theory, become more and more relevant with increasing $N$ and drive the spontaneous chiral symmetry breaking. Theoretically, this amounts to producing a spontaneous gap term in the quasiparticle Hamiltonian: $H_{\text{int}} = H + m\sigma_z$. Depending on the projected sign of $m$ at each spin-valley, a family of five competing broken-symmetry phases were predicted[14,16]: the quantum valley Hall (QVH) state, the quantum anomalous Hall (QAH) state, the layer antiferromagnetic (LAF) state, the quantum spin Hall (QSH) state, and the "ALL" state, as summarized in **Fig. 1a**. They exhibit distinct Hall conductivities, orbital magnetizations, and layer polarizations. When magnetization and/or layer polarization are coupled to electric and/or magnetic fields, a particular state may be favored[8,14].

In the experimental identification of these competing ordered states, we benefit from the high sample cleanliness (see **Fig. S5**) and low dielectric surrounding of freestanding samples, which maximize the interaction effects in r-TLG. **Figure 1b** shows an optical microscopy image of a suspended, dual-gated r-TLG device. **Figure 1c** shows its scattering scanning near-field optical microscopy image, which confirms the stacking order homogeneity of the sample (see also **Fig. S1**). **Figure 1d-g** show the two-terminal differential conductance $G$ as a function of the top and bottom gate voltages ($V_T$ and $V_B$) and of the electric and magnetic fields ($E$ and $B$).

We first discuss our observations at elevated magnetic fields, where the spontaneously gapped states introduced above adiabatically evolve into symmetry-broken quantum Hall states[28]. Indeed, at $B$ = 3 T, the 12-fold degeneracy of the zeroth Landau level[29] is already fully broken and all plateaus at integer filling factors -6 ≤ $\nu$ ≤ 6 are well resolved (**Fig. 1e**). This underscores the high device quality and the strong influence of exchange interaction in the system. While such symmetry-broken states have been described previously[29,30], we observe surprisingly sharp discontinuities in the conductance within a constant filling factor close to the $E$ = 0 line at $B$ = 3 T (**Fig. 1e**). These discontinuities are present in all the symmetry-broken states with 0 < |ν| < 6 and become prominent at $B$ = 8 T (**Fig. 1f**). Since the top and bottom layers of r-TLG are expected to have different effective couplings to the metallic contacts, such a discontinuous jump in conductivity reflects the layer-polarization of a symmetry-broken state that reverses across the $E$ = 0 line. To investigate this inversion of layer polarization, several measurements sweeping across the layer-polarization transitions at different filling factors and different temperatures but constant charge carrier densities for $B = 8$ T are shown in **Fig. 2a-c** (see also **Fig. S6**). Near the sharp transitions (**Fig. 2a**), an extremely small change in electric field of less than 50 μV/nm determines the spontaneous layer polarization between the two outermost layers, indicating ferroelectric quantum Hall states. This is in stark contrast to the observation in BLG, where similar transitions extend over a broad transition region of several mV/nm with increased conductance[2,7,8,31,32]. The transitions in r-TLG display a pronounced electric hysteresis as shown in the line traces in **Fig. 2b**. This hysteresis vanishes at temperatures above 400 mK (**Fig. 2c**). Unlike the case of AlAs quantum wells[33], we do not find resistive spikes at the first-order quantum phase transitions, consistent with the absence of dissipation: Due to the coincidence of layer and valley degrees of freedom[28], the counterpropagating edge modes between domains of oppositely layer-polarized quantum Hall states are localized at opposite valleys and most likely not fully gapped.

Hereafter we focus on the low-magnetic field regime. At zero magnetic field but high electric fields, in agreement with single-particle theories[14,26,27] and previous experiments[34,35], the opening of a band gap is evidenced by a decrease in conductance, as indicated by the red line trace in **Fig. 1d**. This again confirms the rhombohedral stacking order, since such a band gap opening is impossible for Bernal TLG[25,27,34,36,37]. However, unexpected from the single-particle picture, instead of monotonously increasing with decreasing the electric field, the conductance drops again at small electric fields indicating the presence of a spontaneous energy gap due to the electron-electron interactions[14,16,22]. This is consistent with transport spectroscopy measurements on r-TLG that showed a similar spontaneous gap at vanishing electric fields[38]. To elucidate this spontaneously gapped phase, **Figure 1g** shows the conductance as a function of the electric and magnetic fields at charge neutrality. Clearly, the phase can be suppressed by the electric field but strengthened by the magnetic field. This observation agrees well with theoretical predictions[14,16,22] and previous measurements on rhombohedral graphene few-layers[25,31,38–40], and allows for identifying the emergent phase at vanishing electric fields (Phase I) as a LAF state (or a canted anti-ferromagnetic (CAF) state at finite magnetic fields[15,16,38]). With increasing electric field, the layer-balanced LAF/CAF state becomes unstable towards a fully layer-polarized QVH state (Phase II). The two transition lines between the CAF and QVH states are marked by an increase in conductance and become linear at high magnetic fields. This observation is in harmony with theoretical predictions[22,41,42] and previous measurements[2,5,6,31] on BLG. (Note that the transition lines are very sharp, indicating the absence of an additional intermediate metallic phase proposed in BLG[43] and the absence of multiple domains[31] within the device.) Similar to the case of BLG, signatures for other interaction-driven phases can be observed at low magnetic and electric fields[1,2,8]. As shown in the inset of **Fig. 1g**, indications of a local conductance minimum at a (negative) constant electric field can be observed in all devices. Thus, we examine the remaining spontaneously gapped phases below to shed light onto the yet unknown ground states at vanishing fields and charge carrier densities.

**Figure 3a,b** show the differential conductance as a function of charge carrier density and electric field at $B$ = 0.2 and 0.5 T. Distinct quantum Hall plateaus at $\nu$ = 0, ±3, ±6 together with several fainter plateaus at $\nu$ = -1, -2, -4, -5 are readily identifiable even at these weak magnetic fields. The insulating $\nu$ = 0 state can again be identified as the CAF state (I) that transitions to the two QVH states (II) at high electric fields. In agreement with the previous observations, the critical electric field shifts towards higher electric fields with increasing magnetic field (**Fig. 1g**). The observation of prominent plateaus at $\nu$ = ±6 and especially $\nu$ = ±3 at these low magnetic fields, however, is surprising. Remarkably, while the $\nu$ = ±6 states only show a weak electric field dependence and is visible at zero electric field, the $\nu$ = ±3 states stabilizes at finite electric fields only. To inspect the evolution of these states in magnetic field, fan diagrams down to $B$ = 0 were recorded at various electric fields, as exemplified by **Fig. 3c**.

To enhance the visibility of the $\nu$ = ±3 and ±6 states and examine the electric field dependence, **Figure 3d-f** show the derivatives of the fan diagrams with respect to the charge carrier density (see **Fig. S7** and **Fig. S8** for data from additional devices). At zero electric field (**Fig. 3d**), the $\nu$ = ±3 states are absent whereas the $\nu$ = ±6 states persist down to ~ 100 mT. With increasing electric field, the $\nu$ = ±3 states become more and more distinct and even stabilize at exceptionally low magnetic fields of less than 50 mT. In the vicinity of these states, several distinct lines with the same slopes are visible. Such conductance fluctuations originate from localized states and indicate the presence of an energy gap[1,44]. Overall, compared to the aforementioned theory, the observations align well with the stabilization of the "ALL" (III) and QAH (IV) states that were predicted[14] to have Chern numbers ±3 and ±6, respectively, as featured in **Fig. 1a**. Since they have nontrivial orbital magnetization[14,16], both states are expected to be favored in external magnetic fields. Moreover, as the "ALL" state is partially spin- and, in particular, layer-polarized[14,16], the $\nu$ = ±3 states are expected to be favored in electric fields. Our observations are indeed consistent with these predicted properties.

To shed more light onto the nature of the spontaneous quantum Hall states and to examine their competition for the ground state, the two-terminal conductance at various different filling factors was tracked while sweeping the magnetic field through zero. Since the QAH and "ALL" states both exhibit orbital magnetism[14,16], they are expected to display hysteretic behavior in the longitudinal and Hall conductances[17–19]. The same is true for the two-terminal conduction that contains contributions from both conductances. (While indications of magnetism in rhombohedral graphene have been observed previously, their origins are irrelevant to the QAH and "ALL" states[25,40].) **Figure 4a,c** feature the conductance as functions of magnetic field for both forward and backward sweeps at constant filling factors at $E$ = -20 mV/nm. For comparison, **Figure 4b** shows the conductance at constant charge carrier densities, i.e., varying filling factors. In all cases, forward and backward sweeps are mirror symmetric with respect to $B$ = 0. First, noticeable hysteresis can be evidenced at filling factors close to the $\nu$ = -3 and the $\nu$ = -6 states, while it is absent at filling factors far away, e.g., $\nu$ = 0 and $\nu$ = -8 (**Fig. 4c**). Second, there is no noticeable hysteresis at any fixed charge carrier density (**Fig. 4b**). These hysteretic features are consistent with the stabilization of the QAH state at $\nu$ = ±6 and the "ALL" state at $\nu$ = ±3. However, the fact that their conductances drop to values below 1 $e^2/h$ at $B$ = 0 implies that there are no percolating edge states at $B$ = 0, and that both states lose the competition against the CAF or QVH state at $B$ = 0. Besides, it has also been pointed out for similar devices that the current annealing induced disorder close to the electrical contacts could hinder the coupling between edge channels and electrical contacts, thereby obscuring the observation of conductance at $B = 0$[45]. In this scenario, the magnetic hysteresis supported QAH state near zero electric field and "ALL" state at elevated electric fields might persist down to $B = 0$, although they cannot be accessed via two-terminal transport measurements.

To conclude, we have conducted transport measurements on suspended dual-gated r-TLG devices, where Coulomb interaction and electromagnetic tunability are maximized, and discovered a remarkable phase diagram of spontaneously broken-symmetry states near charge neutrality. The layer-balanced insulating LAF/CAF state is observed at low electric fields and transitions to the fully layer-polarized QVH state at high electric fields. At zero electric field, the quantum Hall ferromagnets of the zeroth Landau level are ferroelectrics with spontaneous layer polarizations, as evidenced by the electric hysteresis behavior. Because of their spontaneous orbital magnetizations, as revealed by the magnetic hysteresis behavior, the predicted QAH and "ALL" competing ground states with Chern numbers ±3 and ±6 can be stabilized at vanishing but finite magnetic fields. While future study using multi-terminal devices with still higher quality is necessary to unveil the fate of the QAH, "ALL", and ferroelectric states at strictly $B = 0$, our findings underscore that the natural rhombohedral graphene systems are fertile grounds for strongly interacting electron physics with no need of the delicate moiré engineering[27].

**Materials and Methods**

*Sample Preparation.* r-TLG flakes were exfoliated onto a Si wafer with a 300 nm $SiO_2$ layer and identified by optical microscopy, Raman spectroscopy and atomic force microscopy (AFM). Additionally, scattering scanning near-field optical microscopy (s-SNOM) was employed to ensure stacking order homogeneity down to nanometer scale and to confirm the absence of any structural (ABC-ABA) domain walls[46]. Suspended dual-gated structures were fabricated as follows. First, regions with homogenous stacking order were cut out using standard electron-beam lithography together with reactive-ion etching to prevent a rhombohedral to Bernal stacking transition[46]. Electrical contacts (5/100 nm Cr/Au), a 140 nm SiO₂ spacer and the top gate (5/160 nm Cr/Au) were fabricated consecutively using electron-beam lithography. To decrease the contact resistance, contacts were treated in a UV/Ozone environment for 1 min prior to metal evaporation. Subsequently, samples were submersed in buffered hydrofluoric acid to remove the SiO₂ spacer and 150 nm of the SiO₂ below the R-TLG flake. The devices were transferred to ethanol and dried in a critical point dryer to prevent collapse of the suspended graphene devices. A schematic of a suspended device is shown in **Fig. S2**.

*Quantum transport.* All measurements were performed in a dilution refrigerator at temperatures below 10 mK (unless specified otherwise) using a standard lock-in technique at an AC frequency of 78 Hz and currents below 5 nA. Prior to any measurement, the devices were cleaned in-situ via current annealing (see **Fig. S3** for details). By adjusting the voltages $V_\mathrm{B}$ and $V_\mathrm{T}$ of the silicon back gate and the gold top gate, respectively, both the charge carrier density $n = C_\mathrm{B}(\alpha V_\mathrm{T} + V_\mathrm{B})/e$ and the electric field $E = C_\mathrm{B}(\alpha V_\mathrm{T} - V_\mathrm{B})/2\varepsilon_0$ could be tuned independently[2]. Here, $C_\mathrm{B}$ is the capacitance per unit area of the bottom gate, $\alpha$ is the ratio of top and bottom gate capacitances $\alpha = C_\mathrm{T}/C_\mathrm{B}$, $e$ is the electron charge, and $\varepsilon_0$ is the vacuum permittivity. The calibration procedure is outlined in the Supplementary Information and in **Fig. S4**. The contact resistance $R_\mathrm{C}$ and the bottom gate capacitance were extracted from the quantum Hall plateaus at $B = 3$ T and various electric fields.

**Acknowledgements**

F.W., F.R.G., N. F. and R.T.W. acknowledge funding from the Center for Nanoscience (CeNS) and by the Deutsche Forschungsgemeinschaft (DFG, German Research Foundation) under Germany's Excellence Strategy-EXC-2111-390814868 (MCQST). R.T.W. acknowledges partial funding from the DFG SPP 2244 (2DMP). F.Z. acknowledges support from the US National Science Foundation under grant numbers DMR-1945351 through the CAREER program and DMR-2105139 through the CMP program.

**Data availability**

All data discussed in the manuscript is included in the main figures or the SI. The data are available from the authors upon reasonable request.

**Competing interests**

The authors have no competing interests.

**Author contributions**

F.W, F.R.G., and N.F. fabricated the samples, F.R.G. performed the SNOM measurements, F.W. and N.F. conducted the electrical measurements with the help of F.R.G. All authors discussed and interpreted the data. The manuscript was written by F.W, F.Z, and R.T.W. with input from all authors.

## References

1. Martin, J., Feldman, B. E., Weitz, R. T., Allen, M. T. & Yacoby, A. Local compressibility measurements of correlated states in suspended bilayer graphene. *Phys. Rev. Lett.* **105,** 256806; 10.1103/PhysRevLett.105.256806 (2010).

2. Weitz, R. T., Allen, M. T., Feldman, B. E., Martin, J. & Yacoby, A. Broken-symmetry states in doubly gated suspended bilayer graphene. *Science* **330,** 812–816; 10.1126/science.1194988 (2010).

3. Mayorov, A. S. *et al.* Interaction-driven spectrum reconstruction in bilayer graphene. *Science* **333,** 860–863; 10.1126/science.1208683 (2011).

4. Bao, W. *et al.* Evidence for a spontaneous gapped state in ultraclean bilayer graphene. *Proc. Natl. Acad. Sci. USA* **109,** 10802–10805; 10.1073/pnas.1205978109 (2012).

5. Velasco, J. *et al.* Transport spectroscopy of symmetry-broken insulating states in bilayer graphene. *Nat. Nanotech.* **7,** 156–160; 10.1038/nnano.2011.251 (2012).

6. Freitag, F., Weiss, M., Maurand, R., Trbovic, J. & Schönenberger, C. Spin symmetry of the bilayer graphene ground state. *Phys. Rev. B* **87,** 161402; 10.1103/PhysRevB.87.161402 (2013).

7. Velasco, J. *et al.* Competing ordered states with filling factor two in bilayer graphene. *Nat. Commun.* **5,** 4550; 10.1038/ncomms5550 (2014).

8. Geisenhof, F. R. *et al.* Quantum anomalous Hall octet driven by orbital magnetism in bilayer graphene. *Nature* **598,** 53–58; 10.1038/s41586-021-03849-w (2021).

9. Zhou, H. *et al.* Isospin magnetism and spin-polarized superconductivity in Bernal bilayer graphene. *Science* **375,** 774–778; 10.1126/science.abm8386 (2022).

10. La Barrera, S. C. de *et al.* Cascade of isospin phase transitions in Bernal-stacked bilayer graphene at zero magnetic field. *Nat. Phys.* **18,** 771–775; 10.1038/s41567-022-01616-w (2022).

11. Seiler, A. M. *et al.* Quantum cascade of correlated phases in trigonally warped bilayer graphene. *Nature* **608,** 298–302; 10.1038/s41586-022-04937-1 (2022).

12. Zhang, Y. *et al.* Spin-Orbit Enhanced Superconductivity in Bernal Bilayer Graphene. *arXiv:2205.05087v1* (2022).

13. Zhang, F., Min, H. & MacDonald, A. H. Competing ordered states in bilayer graphene. *Phys. Rev. B* **86,** 155128; 10.1103/PhysRevB.86.155128 (2012).

14. Zhang, F., Jung, J., Fiete, G. A., Niu, Q. & MacDonald, A. H. Spontaneous quantum Hall states in chirally stacked few-layer graphene systems. *Phys. Rev. Lett.* **106,** 156801; 10.1103/PhysRevLett.106.156801 (2011).

15. Zhang, F. & MacDonald, A. H. Distinguishing spontaneous quantum Hall states in bilayer graphene. *Phys. Rev. Lett.* **108,** 186804; 10.1103/PhysRevLett.108.186804 (2012).

16. Zhang, F. Spontaneous chiral symmetry breaking in bilayer graphene. *Synth. Met.* **210,** 9–18; 10.1016/j.synthmet.2015.07.028 (2015).

17. Sharpe, A. L. *et al.* Emergent ferromagnetism near three-quarters filling in twisted bilayer graphene. *Science* **365,** 605–608; 10.1126/science.aaw3780 (2019).

18. Serlin, M. *et al.* Intrinsic quantized anomalous Hall effect in a moiré heterostructure. *Science* **367,** 900–903; 10.1126/science.aay5533 (2020).

19. Polshyn, H. *et al.* Electrical switching of magnetic order in an orbital Chern insulator. *Nature* **588,** 66–70; 10.1038/s41586-020-2963-8 (2020).

20. Chen, G. *et al.* Tunable correlated Chern insulator and ferromagnetism in a moiré superlattice. *Nature* **579,** 56–61; 10.1038/s41586-020-2049-7 (2020).

21. Tschirhart, C. L. *et al.* Imaging orbital ferromagnetism in a moiré Chern insulator. *Science* **372,** 1323–1327; 10.1126/science.abd3190 (2021).

22. Nandkishore, R. & Levitov, L. Flavor Symmetry and Competing Orders in Bilayer Graphene. *arXiv:1002.1966* (2010).

23. Zhou, H. *et al.* Half- and quarter-metals in rhombohedral trilayer graphene. *Nature* **598,** 429–433; 10.1038/s41586-021-03938-w (2021).

24. Zhou, H., Xie, T., Taniguchi, T., Watanabe, K. & Young, A. F. Superconductivity in rhombohedral trilayer graphene. *Nature* **598,** 434–438; 10.1038/s41586-021-03926-0 (2021).

25. Shi, Y. *et al.* Electronic phase separation in multilayer rhombohedral graphite. *Nature* **584,** 210–214; 10.1038/s41586-020-2568-2 (2020).

26. Koshino, M. & McCann, E. Trigonal warping and Berry's phase Nπ in ABC-stacked multilayer graphene. *Phys. Rev. B* **80,** 165409; 10.1103/PhysRevB.80.165409 (2009).

27. Zhang, F., Sahu, B., Min, H. & MacDonald, A. H. Band structure of ABC -stacked graphene trilayers. *Phys. Rev. B* **82,** 35409; 10.1103/PhysRevB.82.035409 (2010).

28. Zhang, F., Tilahun, D. & MacDonald, A. H. Hund's rules for the N=0 Landau levels of trilayer graphene. *Phys. Rev. B* **85**; 10.1103/PhysRevB.85.165139 (2012).

29. Lee, Y. *et al.* Multicomponent Quantum Hall Ferromagnetism and Landau Level Crossing in Rhombohedral Trilayer Graphene. *Nano Lett.* **16,** 227–231; 10.1021/acs.nanolett.5b03574 (2016).

30. Zhang, L., Zhang, Y., Camacho, J., Khodas, M. & Zaliznyak, I. The experimental observation of quantum Hall effect of l=3 chiral quasiparticles in trilayer graphene. *Nat. Phys.* **7,** 953–957; 10.1038/nphys2104 (2011).

31. Geisenhof, F. R. *et al.* Impact of Electric Field Disorder on Broken-Symmetry States in Ultraclean Bilayer Graphene. *Nano Lett.* **22,** 7378–7385; 10.1021/acs.nanolett.2c02119 (2022).

32. Pan, C. *et al.* Layer Polarizability and Easy-Axis Quantum Hall Ferromagnetism in Bilayer Graphene. *Nano Lett.* **17,** 3416–3420; 10.1021/acs.nanolett.7b00197 (2017).

33. Poortere, E. P. de, Tutuc, E., Papadakis, S. J. & Shayegan, M. Resistance spikes at transitions between quantum hall ferromagnets. *Science* **290,** 1546–1549; 10.1126/science.290.5496.1546 (2000).

34. Zou, K., Zhang, F., Clapp, C., MacDonald, A. H. & Zhu, J. Transport studies of dual-gated ABC and ABA trilayer graphene. Band gap opening and band structure tuning in very large perpendicular electric fields. *Nano Lett.* **13,** 369–373; 10.1021/nl303375a (2013).

35. Lui, C. H., Li, Z., Mak, K. F., Cappelluti, E. & Heinz, T. F. Observation of an electrically tunable band gap in trilayer graphene. *Nat. Phys.* **7,** 944–947; 10.1038/nphys2102 (2011).

36. Aoki, M. & Amawashi, H. Dependence of band structures on stacking and field in layered graphene. *Solid State Commun.* **142,** 123–127; 10.1016/j.ssc.2007.02.013 (2007).

37. Koshino, M. Interlayer screening effect in graphene multilayers with ABA and ABC stacking. *Phys. Rev. B* **81,** 125304; 10.1103/PhysRevB.81.125304 (2010).

38. Lee, Y. *et al.* Competition between spontaneous symmetry breaking and single-particle gaps in trilayer graphene. *Nat. Commun.* **5,** 5656; 10.1038/ncomms6656 (2014).

39. Myhro, K. *et al.* Large tunable intrinsic gap in rhombohedral-stacked tetralayer graphene at half filling. *2D Mater.* **5,** 45013; 10.1088/2053-1583/aad2f2 (2018).

40. Lee, Y. *et al.* Gate Tunable Magnetism and Giant Magnetoresistance in ABC-stacked Few-Layer Graphene. *arXiv:1911.04450* (2019).

41. Kharitonov, M. Canted antiferromagnetic phase of the ν=0 quantum Hall state in bilayer graphene. *Phys. Rev. Lett.* **109,** 46803; 10.1103/PhysRevLett.109.046803 (2012).

42. Kharitonov, M. Antiferromagnetic state in bilayer graphene. *Phys. Rev. B* **86,** 195435; 10.1103/PhysRevB.86.195435 (2012).

43. Li, J. *et al.* Metallic Phase and Temperature Dependence of the ν=0 Quantum Hall State in Bilayer Graphene. *Phys. Rev. Lett.* **122,** 97701; 10.1103/PhysRevLett.122.097701 (2019).

44. Lee, D. S., Skákalová, V., Weitz, R. T., Klitzing, K. von & Smet, J. H. Transconductance fluctuations as a probe for interaction-induced quantum Hall states in graphene. *Phys. Rev. Lett.* **109,** 56602; 10.1103/PhysRevLett.109.056602 (2012).

45. Freitag, F., Weiss, M., Maurand, R., Trbovic, J. & Schönenberger, C. Homogeneity of bilayer graphene. *Solid State Commun.* **152,** 2053–2057; 10.1016/j.ssc.2012.09.001 (2012).

46. Geisenhof, F. R. *et al.* Anisotropic Strain-Induced Soliton Movement Changes Stacking Order and Band Structure of Graphene Multilayers. Implications for Charge Transport. *ACS Appl. Nano Mater.* **2,** 6067–6075; 10.1021/acsanm.9b01603 (2019).

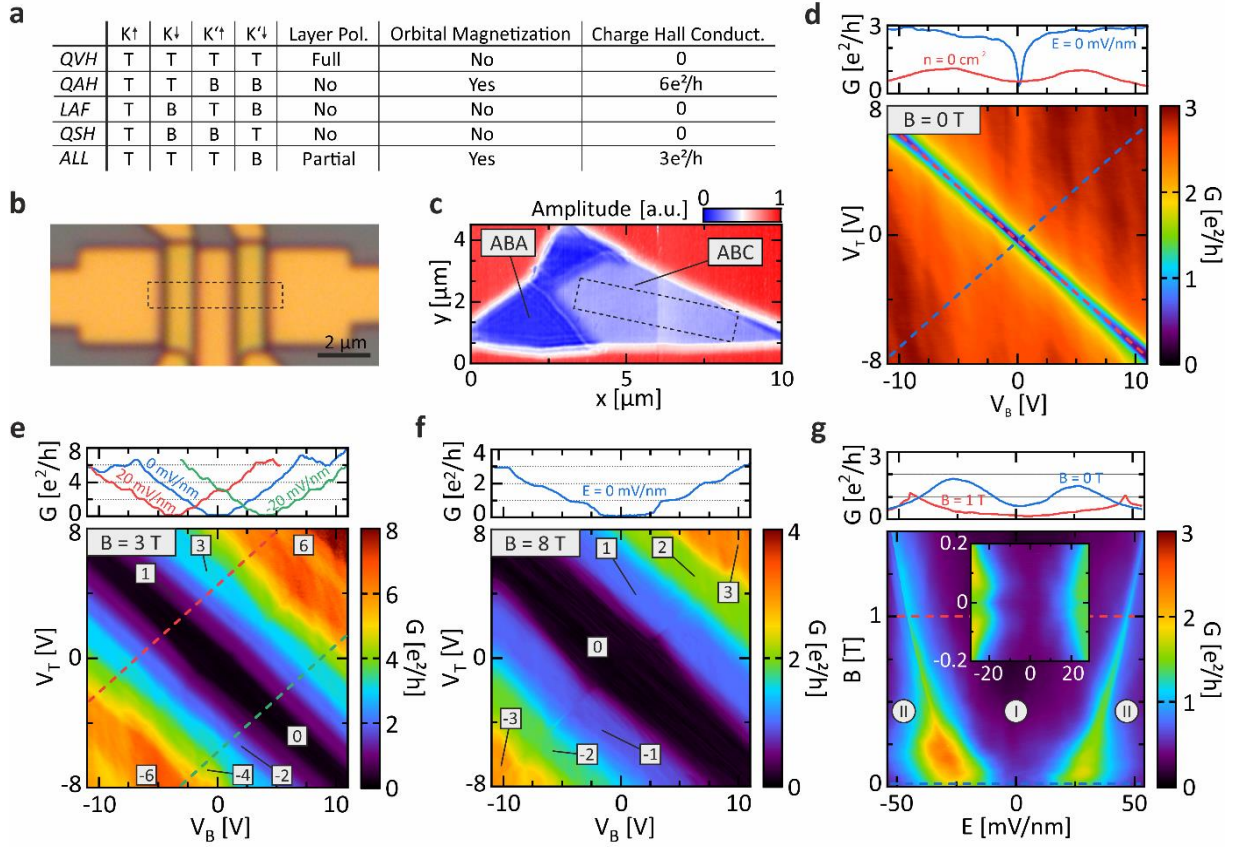

**Fig. 1 | Microscopy and transport measurements. a,** Overview of the five predicted spontaneous quantum Hall states in r-TLG with their corresponding layer polarizations (T for top and B for bottom), total orbital magnetizations, and charge Hall conductivities. **b-c** Optical microscopy image in **b** of a suspended dual-gated r-TLG sample (black dotted box) together with the corresponding s-SNOM image in **c**. The vertical discontinuity in **c** in the ABC region is a measurement artifact. **d-f,** Differential conductance map as a function of top and bottom gate voltages at $B = 0$ T in **d**, $B = 3$ T in **e**, and $B = 8$ T in **f**. Landau level filling factors are indicated by numerals. Line cuts (dotted lines in the contour plots) at constant charge carrier density $n$ or electric field $E$ are shown in the top panel. The dotted line at vanishing electric field is omitted in **e** and **f** for better visibility of the discontinuity in the conductance. **g,** Differential conductance map as a function of $E$ and $B$ at charge neutrality. Line cuts at constant $B$ are shown in the top panel. The inset shows a zoom-in close to vanishing fields. The low conductance regions can be identified as the LAF/CAF (I) and QVH (II) phases.

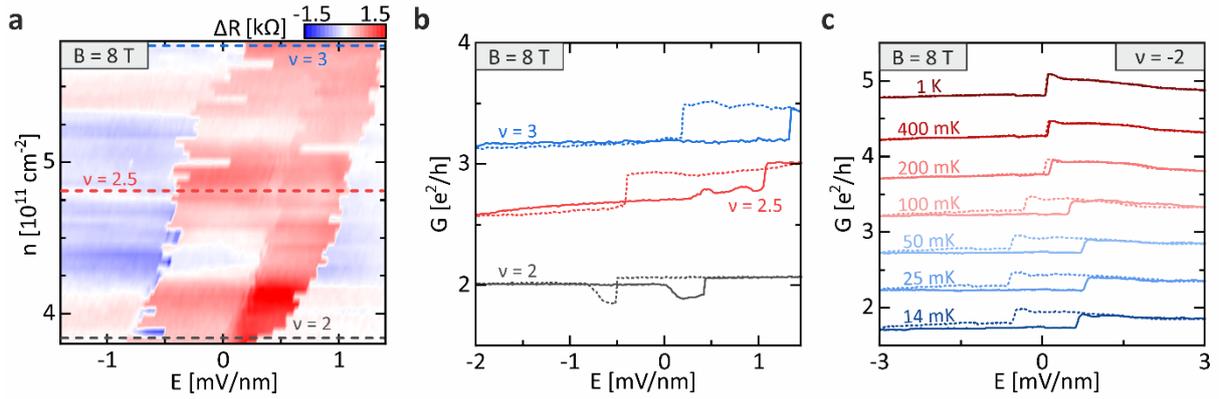

**Fig. 2 | Electric hysteresis of quantum Hall ferromagnets. a**, Map of the resistance difference $\Delta R$ between forward and backward sweeps across the transition line shown in Fig. 1f as a function of electric field $E$ and charge carrier density $n$ at $B = 8$ T. **b**, Line cuts of forward (solid) and backward (dotted) sweeps at filling factors $\nu = 2$, 2.5 and 3 as shown in **a**. **c**, Forward (solid) and backward (dotted) sweeps at $\nu = -2$ and $B = 8$ T for different temperatures between 14 mK (blue) and 1 K (red). The lines are offset by 0.5 $e^2/h$ with respect to each other for clarity.

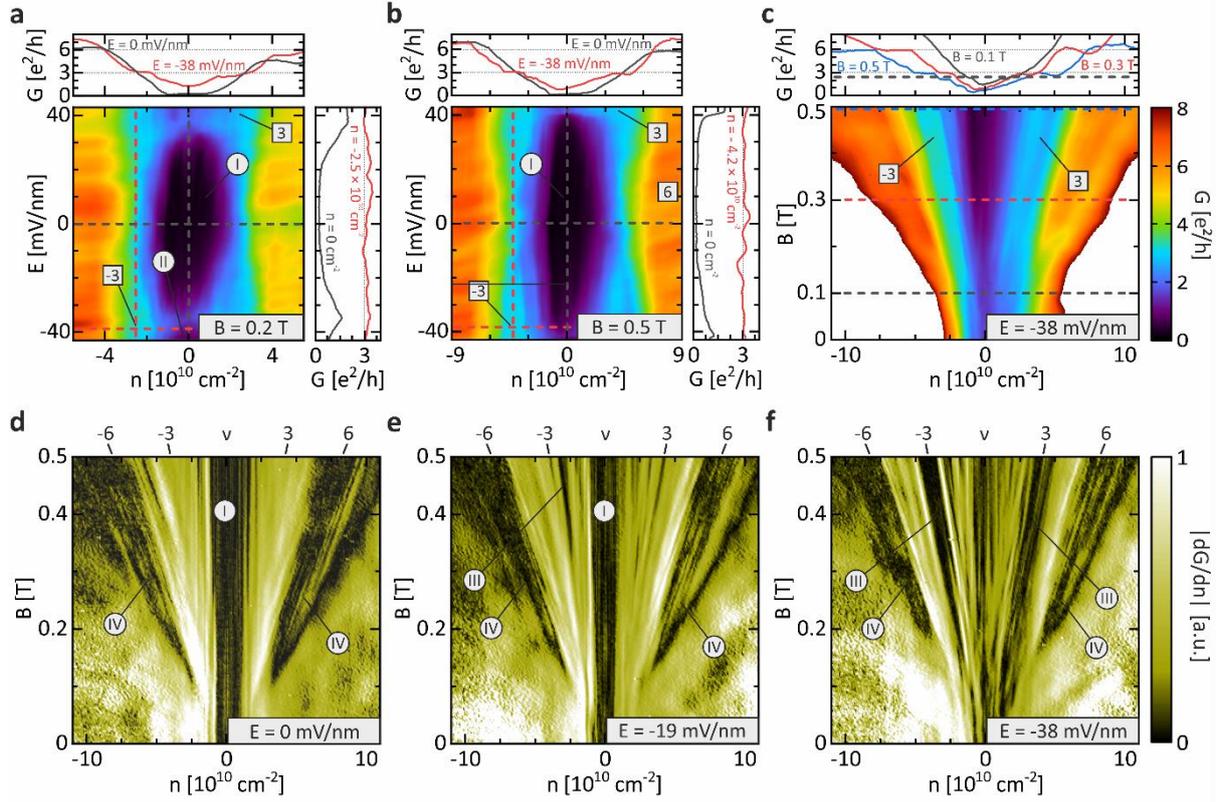

**Fig. 3 | Anomalous quantum Hall states at low magnetic fields. a-b**, Differential conductance map as a function of electric field $E$ and charge carrier density $n$ at $B = 0.2$ T in **a** and $B = 0.5$ T in **b**. The top and right panels show the conductances along lines of constant $E$ and constant $n$, respectively, as indicated by the dotted lines in the main panels. **c**, Fan diagram of the differential conductance at $E = -38$ mV/nm. The top panel shows the conductance along lines of constant $B$, as indicated by the dotted lines in the main panel. **d-f**, Fan diagrams of the derivative of the conductance with respect to $n$ at $E = 0$ mV/nm in **d**, $E = -19$ mV/nm in **e**, and $E = -38$ mV/nm in **f**. The filling factors and their corresponding slopes are indicated on the top of each panel. The roman numerals indicate the predicted spontaneous quantum Hall states, namely, the LAF/CAF (I), QVH (II), ALL (III), and QAH (IV) states.

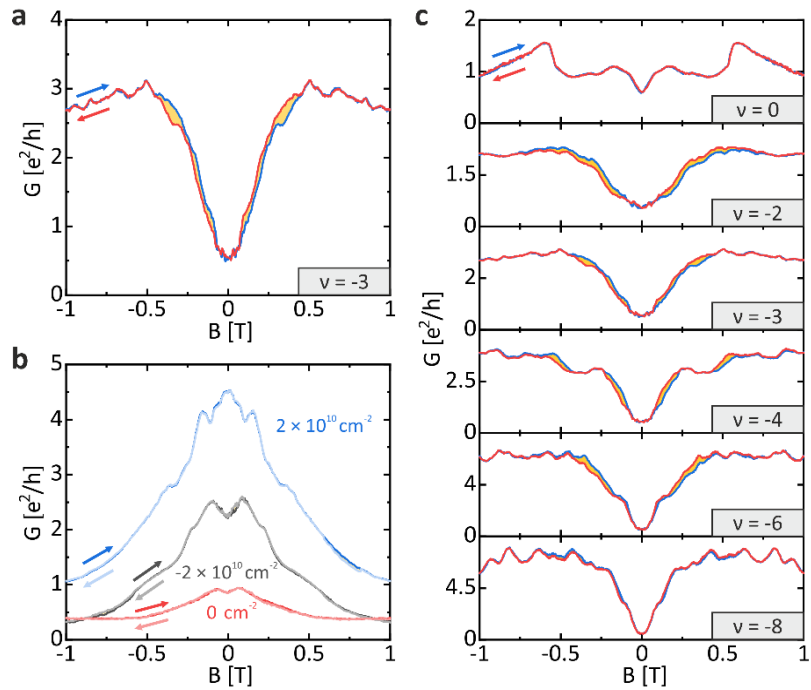

**Figure 4 | Magnetic hysteresis at a constant electric field $E$ = -20 mV/nm. a**, Conductance as a function of magnetic field $B$ at the filling factor $\nu$ = -3. The hysteresis loop area formed by the forward (blue) and backward (red) sweeps is shaded in yellow. **b**, Conductance as a function of $B$ at the densities $n$ = -2 × $10^{10}$ cm$^{-2}$ (black), $n$ = 0 cm$^{-2}$ (red), and $n$ = 2 × $10^{10}$ cm$^{-2}$ (blue). The forward and backward sweeps are indicated in dark and bright colors, respectively. **c**, The same as **a**, but for a sequence of filling factors ranging from $\nu$ = 0 to $\nu$ = -8.

Supplementary Information:

# Ferroelectric and anomalous quantum Hall states
# in bare rhombohedral trilayer graphene

Felix Winterer[1], Fabian R. Geisenhof[1], Noelia Fernandez[1,2],

Anna M. Seiler[1,2], Fan Zhang[3], R. Thomas Weitz[1,2,*]

[1] Physics of Nanosystems, Department of Physics, Ludwig-Maximilians-Universität München, Amalienstrasse 54, Munich 80799, Germany

[2] 1st Institute of Physics, Faculty of Physics, University of Göttingen, Friedrich-Hund-Platz 1, Göttingen 37077, Germany

[3] Department of Physics, University of Texas at Dallas, Richardson, TX, 75080, USA

*Corresponding author. Email: thomas.weitz@uni-goettingen.de

## Identification and Characterization of Suitable ABC-Stacked Trilayer Graphene Flakes

In a first step, r-TLG is located and identified via optical microscopy. **Figure S1a** shows a microscope image of the sample presented in the main text. In order to reveal the stacking order as well as stacking faults and domain walls, a combination of Raman spectroscopy and scattering scanning near-field optical microscopy (s-SNOM)[1] has been employed. **Figure S1b,c** shows maps of the FWHM of the 2D Raman peak and the s-SNOM amplitude, respectively. Both maps exhibit two distinct regions with low and high FWHM (Raman) as well as amplitude (s-SNOM). **Figure S1d** shows the Raman spectra corresponding to these two regions. Comparing the Raman spectra to previous calibration data and literature allows to identify the regions with Bernal (ABA) and rhombohedral (ABC) stacking order[1,2]. By matching the different domains in the Raman map to the regions of different amplitude in the s-SNOM image, the stacking domains can be resolved down with nanometer resolution confirming the absence of stacking faults and domain walls[1]. In order to prevent transformation of stacking order, ribbons with homogeneous ABC stacking order have been etched out using reactive ion etching prior to defining electrical contacts[1].

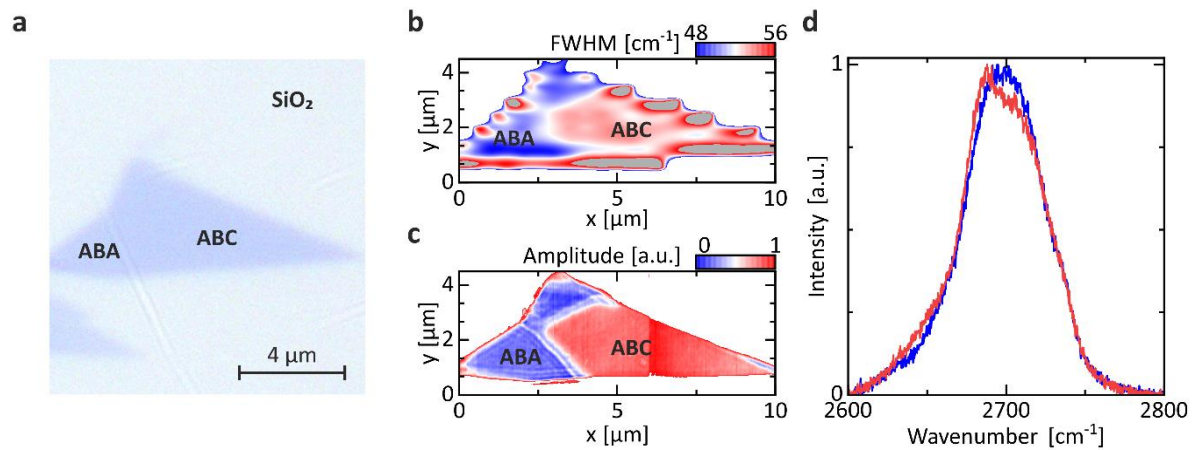

**Figure S1 Stacking Order Characterization: a,** Optical microscope image of the trilayer graphene flake presented in the main text. **b-c,** Map of the FWHM of the 2D Raman mode in **b** and map of the s-SNOM amplitude in **c** of the trilayer graphene flake shown in **a**. Two regions with different 2D-peak FWHMs and s-SNOM amplitudes are readily visible and can be identified with ABA (blue) and ABC (red) stacking order. **d,** Single Raman spectra of the 2D mode taken at a spot inside the ABA (blue) and ABC (red) regions as shown in **b,c**.

## Device Schematic

A schematic cross-section and 3d-rendering of a suspended, dual-gated sample is shown in **Figure S2**.

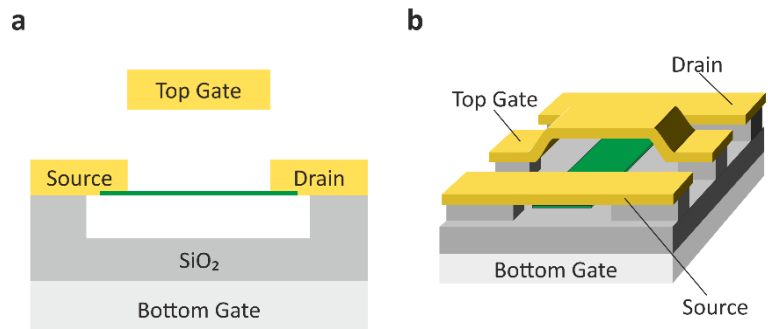

**Figure S2 Device Geometry:** Line cut along the graphene axis in **a** and 3D view in **b** of a dual-gated graphene device with silicon bottom gate and a gold top gate.

## Current Annealing

Suspended graphene devices typically have a substantial amount of contaminants and residues on the graphene surface that greatly reduce the quality of the device. Current annealing is a very effective method to remove these contaminants in-situ by forcing a large current through the graphene device[3,4]. Due to dissipation, the high current density is accompanied by strong heating that removes the absorbed contaminations[3,4]. **Figure S3a** shows top gate voltage sweeps illustrating the increase in device quality during annealing. After current annealing, the resistance peak associated with the charge neutrality point of trilayer graphene is clearly visible. Current annealing was implemented as follows. The drain voltage $V_{BIAS}$ across the graphene was ramped in consecutive runs to higher and higher values. Between each ramp, the quality of the device was monitored via gate sweeps. At some voltage, the drain current starts to saturate and the overall device resistance starts to increase (see **Figure S3b,c**). The best results were achieved, when annealing up to drain voltages approximately 0.7 – 1.5 V above the onset of the resistance increase.

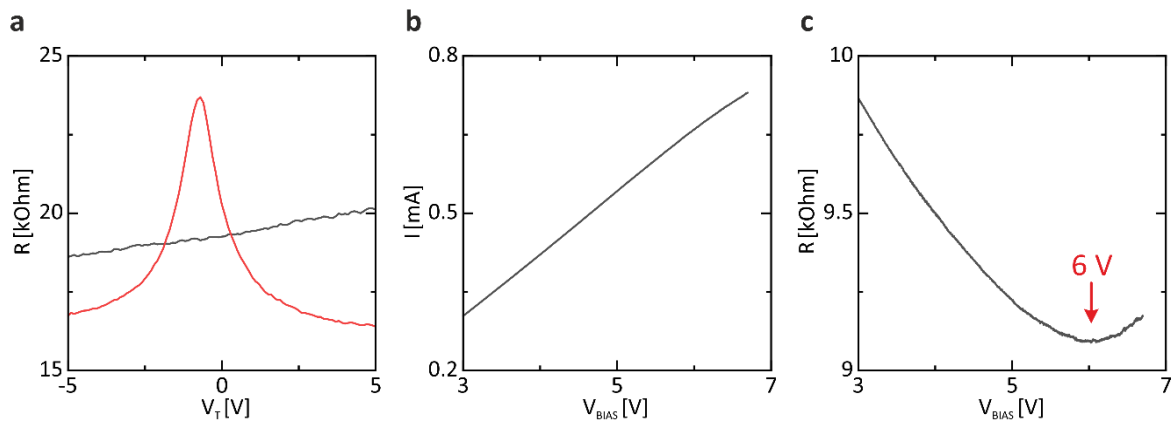

**Figure S3 Current Annealing: a,** Two-terminal resistance (including the line resistance of the cryostat) as function of the top gate voltage before (black) and after (red) current annealing. **b-c,** Drain current and resistance of the final drain-source voltage ramp during the current annealing procedure. The voltage bias was ramped up to 0.7 V above the onset of the resistance increase at approximately 6 V (red arrow). The curve shown in **a** was measured directly after this ramp.

## Calibration of Dual Gate Transport

In dual-gated samples, the charge carrier density $n = C_B(\alpha V_T + V_B)/e$ and the electric field $D = C_B(\alpha V_T - V_B)/2\varepsilon_0$ can be tuned independently as a function of bottom gate voltage $V_B$ and top gate voltage $V_T$. Here, $C_B$ is the capacitance per unit area of the bottom gate and $\alpha = C_T/C_B$ is the ratio between the capacitances of bottom and top gate. Inverting these equations yields $V_B = (en - 2\varepsilon_0 D)/2C_B + V_{B,0}$ and $V_T = (en + 2\varepsilon_0 D)/2\alpha C_B + V_{T,0}$, where the voltage offsets $V_{B,0}$ and $V_{T,0}$ have been introduced to account for residual contaminants that shift the charge neutrality. In order to determine $\alpha$, the conductance of the device is mapped as a function of the bottom gate voltage and top gate voltage at zero magnetic field as shown in **Figure S4a**. In every line of constant top gate voltage, the center of the conductance dip is determined via a Gaussian fit. Collecting all peak positions and performing a linear fit directly yields a value for $\alpha$ as the inverse of the slope. This line fit is essentially the charge neutrality line and moving along this line translates to changing the electric field. The offsets $V_{B,0}$ and $V_{T,0}$ are determined iteratively as follows. First, the conductance is measured as a function of charge carrier density while setting the electric field and magnetic field to zero (see **Figure S4b**). The offsets $V_{B,0}$ and $V_{T,0}$ are selected in order to correct for the displacement of the Gaussian fit from zero charge carrier density. Consecutively, the conductance is measured as a function of electric field at charge neutrality and $B = 0.4$ T (see **Figure S4c**). Here, the offsets $V_{B,0}$ and $V_{T,0}$ are corrected such that the mean of the two peak positions is at zero field. This alignment has shown to yield the most reliable results and relates to the presence of two distinctive symmetric conductance peaks at non-zero magnetic and electric fields (see **Figure 1g** in the main text). This procedure is repeated with alternating sweeps of the charge carrier density and the electric field until $V_{B,0}$ and $V_{T,0}$ converge. In a final step, the electric field offset is fine-tuned using the conductance jump at vanishing electric field shown in **Figure 2**. The capacitance $C_B$ is approximated by geometrical considerations and then refined by matching the quantum Hall plateaus at 3 T and various electric fields to their expected conductance values.

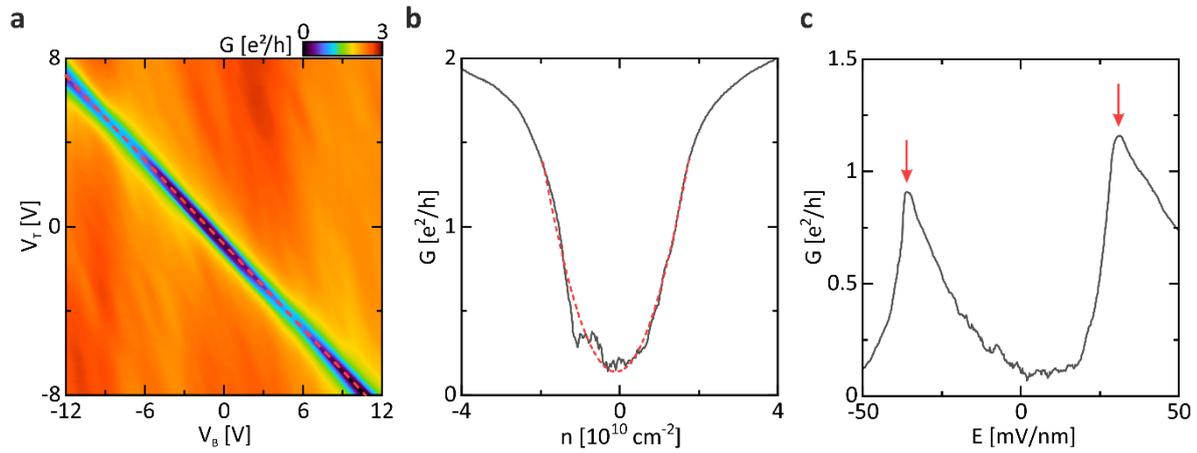

**Figure S4 Calibration of Charge Carrier Density and Electric Field: a,** Conductance as a function of the bottom gate and the top gate voltage. On every column of the plot, the resistance peak position is extracted by performing a Gaussian fit. Connecting all peak positions and fitting a line yields the charge neutrality line (dotted red). **b-c,** The top gate and bottom gate offset voltages are determined iteratively by correcting the center of the Gaussian fit (dotted red) at the zero electric field line (shown in **b**) and by correcting the mean position of the conductance peaks (red arrows) along the charge neutrality line (shown in **c**).

# Charge Carrier Inhomogeneity

A residual charge carrier inhomogeneity of less than $8 \times 10^{-9}$ cm$^{-2}$ is estimated[5] (see **Figure S5)** for the devices shown in the main text. Furthermore, the quantum transport data suggest a low amount of electric field disorder [add here our new nano letters[6].

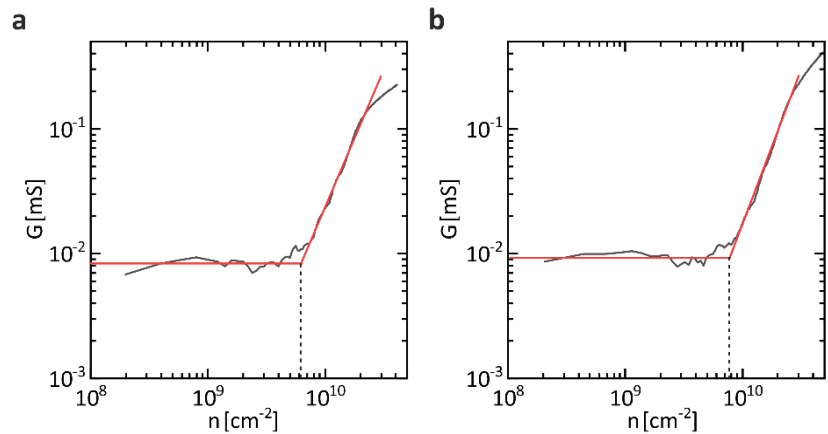

**Figure S5 Charge Carrier Inhomogeneity:** Conductance as a function of charge carrier density at zero perpendicular electric field of the two devices shown in the manuscript and supporting information. The red lines are linear fits with their intersection indicating the residual charge carrier inhomogeneity in the device.

# Inversion of the Layer Polarization

As shown in **Figure 2**, when sweeping from positive to negative electric field, the layer polarization of the electronic state flips. In measurements, this manifests as a sudden jump in the resistance, as bottom and top layer exhibit different couplings to the contact. In addition to **Figure 2, Figure S6** shows maps of the resistance difference between forward and backward sweeps at $B = 3$ T and $B = 8$ T for a larger density range. In both cases, the resistance jump displays a distinct hysteresis. The jump and the hysteresis are readily visible up to a filling factor of $|\nu| = 5$ (n $\approx 3.6 \times 10^{11}$ cm$^{-2}$ for $B = 3$ T) until they start to fade out.

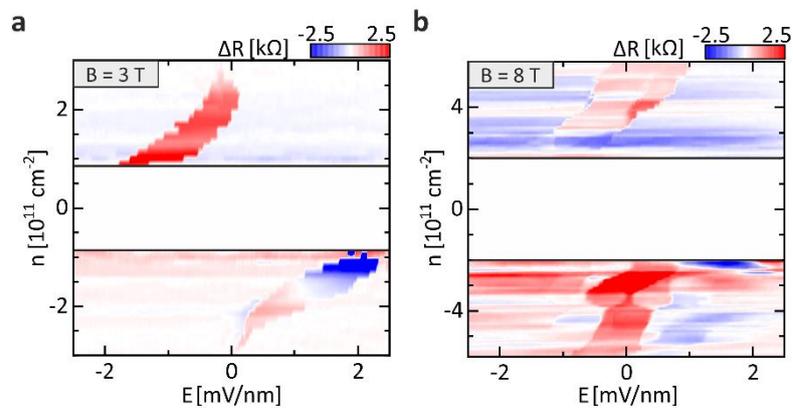

**Figure S6: Inversions of the Layer Polarization:** Map of the resistance difference between forward and backward sweep of the electric field as function of electric field and charge carrier density at B = 3 T in **a** and B = 8 T in **b**. The region with vanishing conductance at zero density has been removed due to large resistance variations to enhance visibility.

## Additional Magnetotransport Data

**Figure S7** shows fan diagrams of an additional device confirming the emergence of conductance plateaus at $\nu = \pm 6$ at vanishing electric field and $\nu = \pm 3$ at high electric field consistent with the appearance of the QAH and the "ALL" state. This is also reproduced in a different sample batch as shown in **Figure S8**.

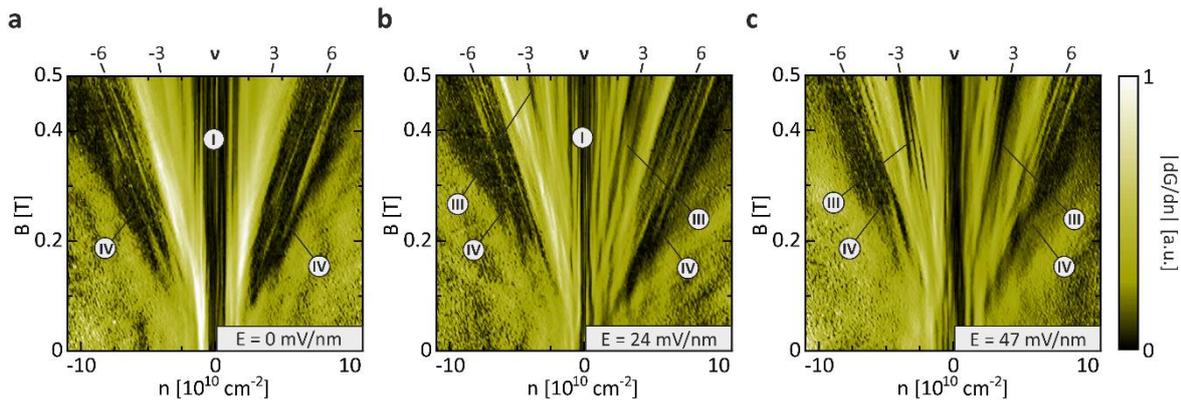

**Figure S7: Additional Data on Quantum Hall States at Low Magnetic Fields:** Fan diagrams of the conductance derivative with respect to the charge carrier density at $E = 0$ mV nm$^{-1}$ in **a**, $E = 24$ mV nm$^{-1}$ in **b** and $E = 47$ mV nm$^{-1}$ in **c**. The filling factors and their corresponding slope are indicated in the top of each panel. The roman numerals indicate the associated spontaneous quantum Hall states, namely the LAF/CAF state (I), the "ALL" state (III) and the QAH state (IV).

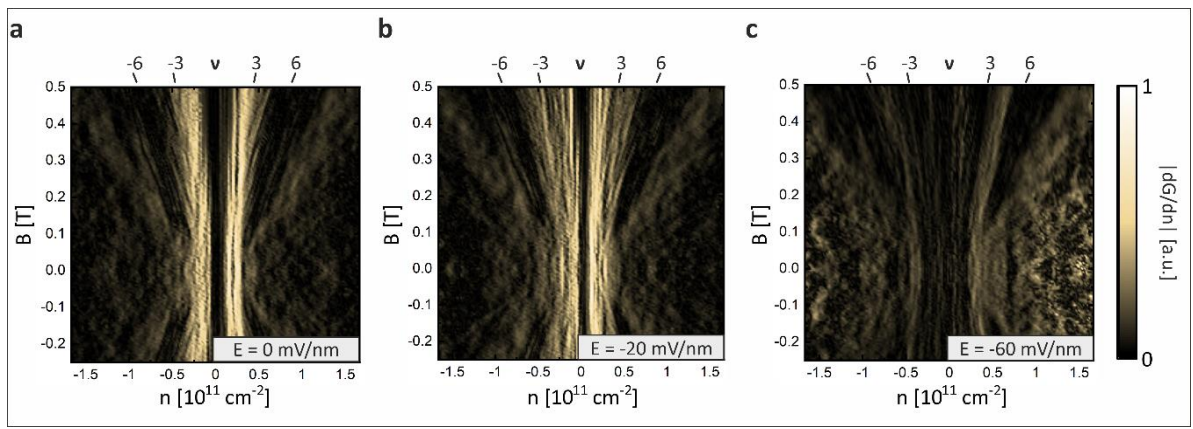

**Figure S8: Magnetotransport Data of an Additional Sample:** Fan diagrams of the conductance derivative with respect to the charge carrier density at $E = 0$ mV nm$^{-1}$ in **a**, $E = -20$ mV nm$^{-1}$ in **b** and $E = -60$ mV nm$^{-1}$ in **c**. The filling factors and their corresponding slope are indicated in the top of each panel.

# References

1. Geisenhof, F. R. *et al.* Anisotropic Strain-Induced Soliton Movement Changes Stacking Order and Band Structure of Graphene Multilayers. Implications for Charge Transport. *ACS Appl. Nano Mater.* **2,** 6067–6075; 10.1021/acsanm.9b01603 (2019).

2. Nguyen, T. A., Lee, J.-U., Yoon, D. & Cheong, H. Excitation Energy Dependent Raman Signatures of ABA- and ABC-stacked Few-layer Graphene. *Sci. Rep.* **4,** 4630; 10.1038/srep04630 (2015).

3. Moser, J., Barreiro, A. & Bachtold, A. Current-induced cleaning of graphene. *Appl. Phys. Lett.* **91,** 163513; 10.1063/1.2789673 (2007).

4. Schedin, F. *et al.* Detection of individual gas molecules adsorbed on graphene. *Nat. Mater.* **6,** 652–655; 10.1038/nmat1967 (2007).

5. Nam, Y., Ki, D.-K., Soler-Delgado, D. & Morpurgo, A. F. Electron–hole collision limited transport in charge-neutral bilayer graphene. *Nat. Phys.* **13,** 1207–1214; 10.1038/nphys4218 (2017).

6. Geisenhof, F. R. *et al.* Impact of Electric Field Disorder on Broken-Symmetry States in Ultraclean Bilayer Graphene. *Nano Lett.* **22,** 7378–7385; 10.1021/acs.nanolett.2c02119 (2022).